# Idea of a new Personality-Type based Recommendation Engine

Animesh Pandey*

November 9, 2013

*Abstract*: Myers-Briggs Type Indicator (MBTI) types depict the psychological preferences by which a person perceives the world and make decisions. There are 4 principal functions through which the people see the world: sensation, intuition, feeling, and thinking. These functions along with the Introverted\Extroverted nature of the person, there are 16 personalities types, the humans are divided into. Here an idea is presented where a user can get recommendations for books, web media content, music and movies on the basis of the users' MBTI type. Only things like books and other media content has been chosen because the preferences in such things are mostly subjective. Apart from the recommended content that is generally generated on the basis of the previous purchases, searches can be enhanced by using the MBTI. A minimalist survey was designed for collecting the data. This has a more than 100 features that show the preference of a personality type. Those include preferences in book genres, music genres, movie genres and even video games genres. After analyzing the data that is collected from the survey, some inferences were drawn from it which can be used to design a new recommendation engine for recommending the content that coincides with the personality of the user.

## 1 Introduction

The World Wide Web (WWW) is a great source of information, which if used intelligently can help the users as well as the businesses. Till now the WWW has undergone many developments. Currently, the applications are running on the Web 2.0 framework and slowly are transitioning to Web 3.0 which is the Semantic Web. Semantic web is the type of WWW from which the information can be yielded using the information present on the web i.e. using the metadata. With this metadata, personalization is done. The world of personalized media continues to enhance itself day to day. Mostly all search engines Google, Bing etc. try to orient results to reflect our social preferences and our past affinities. Facebook generates ads on the basis of your 'liked' pages, or your visits to certain pages. It also prioritizes newsfeed content based on our past engagement. Some major media channels like *The Huffington Post* or *The New York Times* all provide recommended content for users.

According to Econsultancy Report [1] released in June, 2012, majority (52%) of digital marketers say that personalization is fundamental to their online strategy. This raises a question that "How can personalization affect our experiences as consumers and strategies as marketers?" Generally a more personalized web where irrelevant content can be ignored and relevant content can be presented to the user, is welcomed. Changes in customer preference can quickly be recognized by the marketer and the business strategy can be tailored specifically for that customer.

There are a total of 16 Personality types in the Myers-Briggs Type Indicator (MBTI). Each perceives an object differently. If a photograph is shown to a particular type and it has something that his thinking coincides with, then, it will appeal to him before any other object in the photograph. Same can be thought of media content. For example, According to websites like "personalitycafe.com", "typologycentral.com" or "personalitypage.com", INFPs (Introverted Intuitive Feeling Perceiving) might have a propensity towards religious texts or philosophy. Now, if on the internet a person of such type sees a few books where one of the books are of the above mentioned genre, there is higher probability that he might click on it.

*Bachelor of Technology, Information Technology, Jaypee Institute of Information Technology, Noida, Uttar Pradesh, India
Email ID: apanimesh061@gmail.com

## 2 Data Sets used

The data has been collected with an online survey [2] that was hosted at "personalitycafe.com", "typologycentral.com", facebook groups and "reddit.com". After four months a data set of 1020 entries with ratings of over 121 objects was collected.

The dataset had categories of media content like books, video games, music and movies. Each category was divided into genres. Books had 34 Non Fiction genres ad 30 Fiction genres, Music had 25 genres, Movies had 21 genres and Video games had 11 genres.

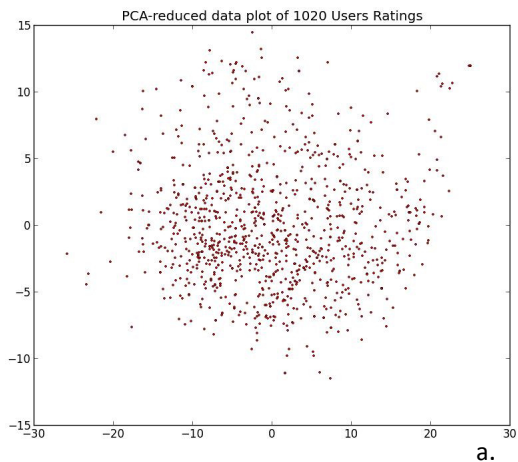
a.

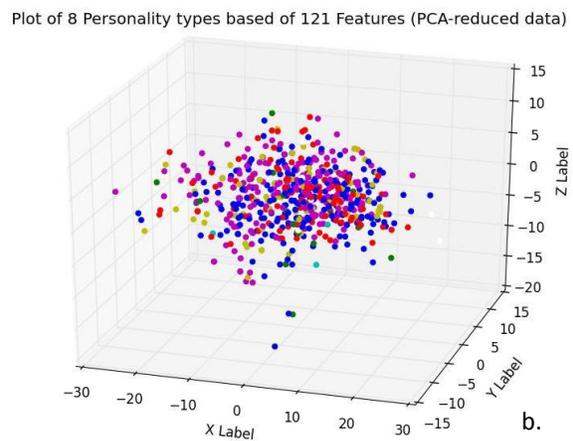
b.

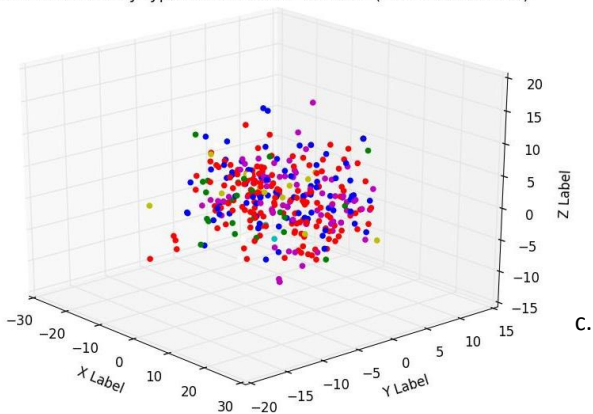
c.

Fig. 1. This figure set is visualization of the data set collected in 2D and 3D.

The first one (a.) is in 2D for all 16 personality types. The other two (b., c.) are for 2 sets of 8 personality types

## 3 Clustering of Data

Now, K-means clustering is performed after applying PCA (Principal Component Analysis) on each media content category. Following figures show the clustering of data into 16 clusters which will show how many Personality types prefer which type of content. There are 1020 data entries where the frequencies of each type were:
'intp': 221, 'intj': 160, 'infj': 134, 'infp': 111, 'istp': 81, 'entp': 76, 'enfp': 71, 'istj': 65, 'isfj': 26, 'isfp': 22, 'entj': 17, 'estp': 12, 'enfj': 11, 'esfp': 5, 'estj': 5 and 'esfj': 3

Performance evaluation of the clusters has been done. Seven evaluation criteria have been considered which are:

a. Homogeneity [3]: Checking whether each cluster contains only members of a single class.
b. Completeness [3]: Checking whether all members of a given class are assigned to the same cluster.
c. V-measure [3, 4] is equivalent to the mutual information normalized by the sum of the label entropies
d. Adjusted Rand index [5, 6, 7] (ARI) is a function that measures the similarity of the two assignments, ignoring permutations and with chance normalization
e. The Mutual Information [5, 6, 7] (AMI) is a function that measures the agreement of the two assignments, ignoring permutations
f. The Silhouette Coefficient [8] is defined for each sample and is composed of two scores:
   - The mean distance between a sample and all other points in the same class.
   - The mean distance between a sample and all other points in the next nearest cluster.

Following table shows the performance of the clustering done on the 1020 data sets collected. Here three methods for clustering have been used. First is the "kmeans++" which selects initial cluster centers for k-mean clustering in a smart way to speed up convergence. Second one is "Random" where we choose k observations (rows) at random from data for the initial centroids and the last one PCA (Principal Component Analysis) based where clustering has been done after reducing the dimension from 1020 to 2. The testing has been done using Python on Windows7 32-bit.

|  | Method | time | homo | compl | v-meas | ARI | AMI | Silhouette |
|---|---|---|---|---|---|---|---|---|
|  | k-means++ | 0.58s | 0.081 | 0.069 | 0.074 | 0.16 | 0.028 | 0.071 |
| Fig. 2.a | Random | 0.40s | 0.077 | 0.067 | 0.071 | 0.161 | 0.025 | 0.053 |
|  | PCA-based | 0.06s | 0.062 | 0.055 | 0.058 | 0.156 | 0.014 | 0.082 |
|  | k-means++ | 0.54s | 0.084 | 0.072 | 0.077 | 0.163 | 0.031 | 0.061 |
| Fig. 2.b | Random | 0.47s | 0.09 | 0.078 | 0.084 | 0.161 | 0.038 | 0.058 |
|  | PCA-based | 0.08s | 0.072 | 0.067 | 0.07 | 0.148 | 0.027 | 0.053 |
|  | k-means++ | 0.50s | 0.064 | 0.055 | 0.059 | 0.157 | 0.013 | 0.045 |
| Fig. 2.c | Random | 0.41s | 0.064 | 0.055 | 0.059 | 0.155 | 0.013 | 0.044 |
|  | PCA-based | 0.06s | 0.058 | 0.053 | 0.056 | 0.147 | 0.012 | 0.034 |
|  | k-means++ | 0.55s | 0.072 | 0.061 | 0.066 | 0.162 | 0.02 | 0.055 |
| Fig. 2.d | Random | 0.38s | 0.069 | 0.058 | 0.063 | 0.164 | 0.017 | 0.064 |
|  | PCA-based | 0.04s | 0.065 | 0.054 | 0.059 | 0.159 | 0.013 | 0.06 |
|  | k-means++ | 0.36s | 0.064 | 0.055 | 0.059 | 0.155 | 0.013 | 0.116 |
| Fig. 2.e | Random | 0.40s | 0.065 | 0.056 | 0.06 | 0.154 | 0.014 | 0.114 |
|  | PCA-based | 0.06s | 0.044 | 0.043 | 0.043 | 0.143 | 0.01 | 0.105 |

Table 1. Performance Evaluation of the Clustering performed

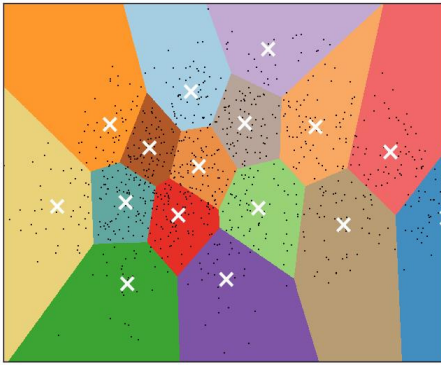

Fig. 2.a. Clustering for Fiction Literature

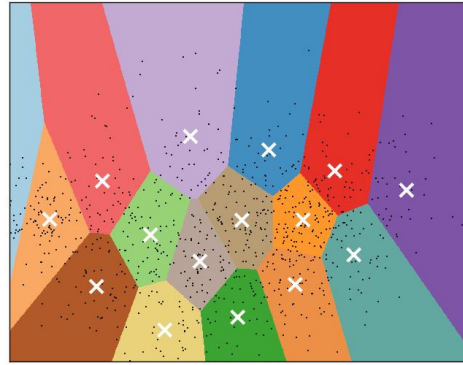

Fig. 2.b. Clustering for Non-Fiction Literature

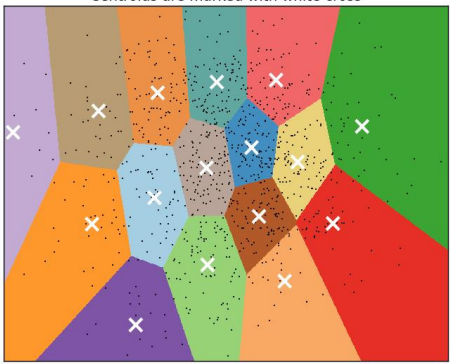

Fig. 2.c. Clustering for Music Genres

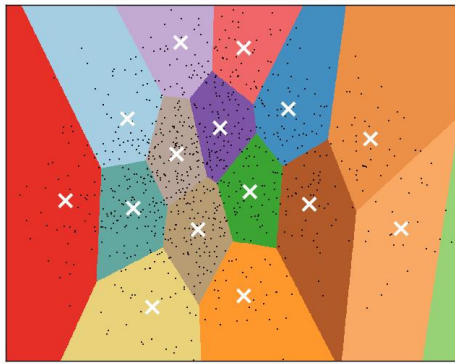

Fig. 2.d. Clustering for Movie Genres

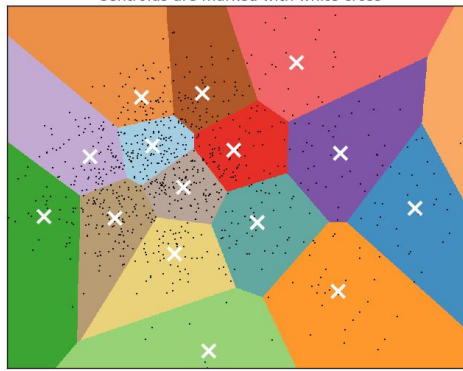

Fig. 2.e. Clustering for Video Games Genres

## 4 Book Genre-Personality Type Relation

Two Non Fiction Book genres namely, Psychology and Religion & Spirituality have been chosen. The comparison is for four personality types: INFP, INTP, INTJ and INFJ. These four have been chosen as they the maximum frequency in the data of 1020 entries.

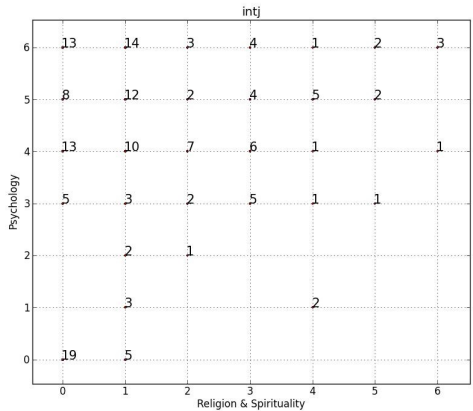
INTJ

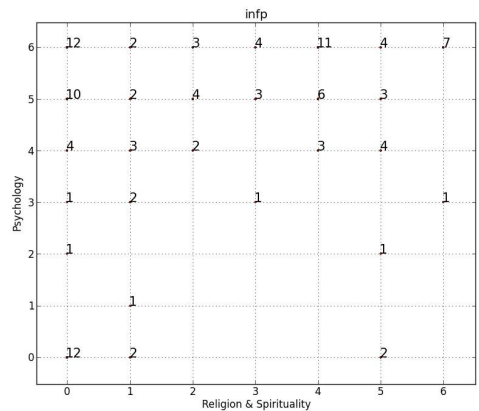
INFP

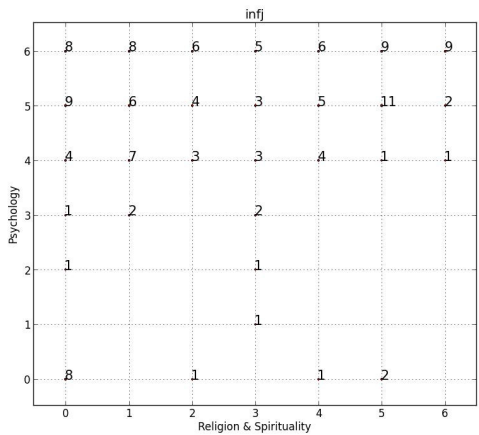
INFJ

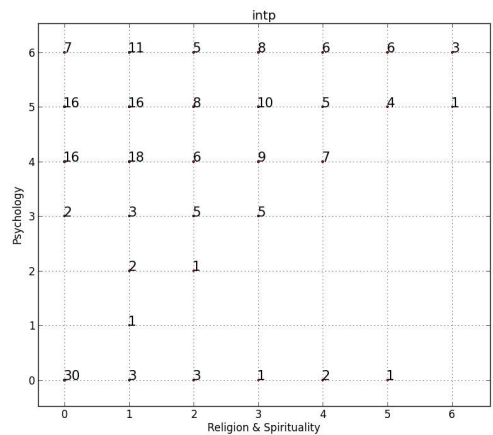
INTP

Fig. 3. Graphs of Ratings vs. Ratings of two book genres

The graphs are between the ratings of the two genres given the users of the four personality types mentioned above. The numbers on the graph present the frequency of the users that gave a particular rating to a genre. The ratings have a range of 0-6 where meaning of each rating is defined below:

I. 0 - No Experience
II. 1 - Dislike strongly
III. 2 - Dislike
IV. 3 - Neutral/No opinion
V. 4 - Mild enjoyment
VI. 5 - Reasonably enjoyable
VII. 6 - Highly enjoyable

# 5 Inferences

Following are some inferences that were deduced from the observations:

1. The data collected is very skewed. The reason could be that as the surveys were hosted on the internet the Extroverts have a lower probability of filling the survey as compared to introverts.

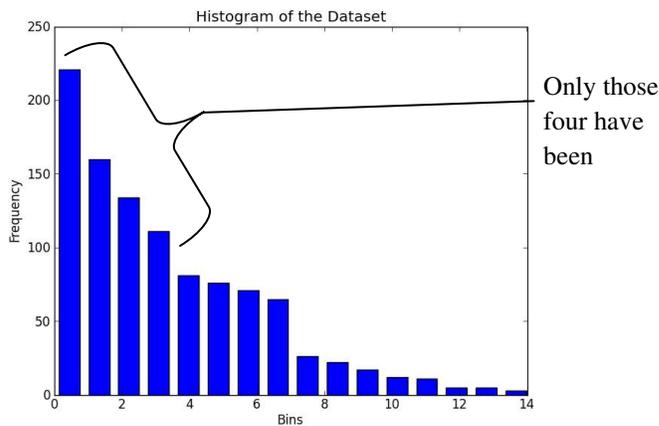

Only those four have been

Fig. 4. The first bar shows INTP, second shows INTJ, third shows INFJ and fourth shows INFP frequency.

2. In Clustering we can see that dividing the set into 16 clusters is not a very good idea at the moment. One reason is that the evaluation is not showing very good results and other is the need of more data.
3. The Book Genre-Personality Type graphs show that the four types are showing higher inclination to Psychology as compared to Religion-Spirituality Genre books. INTP type has most number of users that rated both the genres as '0' which means 'No experience' but many of those who have experience have rated them '4' or '5'.
4. On the basis of visualization (Fig.1.) we can see that there is no special consistency of a particular personality type. A type that has its location somewhere in the middle along with various other types may also have an entry that acts more like an outlier. For example, the personality type denoted by blue dots in Fig. 1.b. has some concentration in the middle as well as there a couple of blue points on extreme corners and near the axes.

# 6 Future Work

More data has to be collected so that it can be checked whether the data will remain skewed or not. Only one pair of book genre has been tested, more pairs need to be tested to better inferences. The media content needs to be analyzed to see which content concedes more with a particular type of personality. This can be done by crawling the web and trying to infer rules from the text that describe a personality type.